\documentclass[sigplan,nonacm]{acmart}
\AtBeginDocument{%
  }

\def\noeditingmarks{}

\usepackage{mfirstuc} %

\usepackage{listings}
\definecolor{codekw}{HTML}{0072B2}
\definecolor{codecmt}{HTML}{767676}
\definecolor{codestr}{HTML}{009E73}
\definecolor{codeapi}{HTML}{CC79A7}

\lstdefinestyle{vkcode}{
  basicstyle=\scriptsize\ttfamily,
  keywordstyle=\bfseries\color{codekw},
  commentstyle=\itshape\color{codecmt},
  stringstyle=\color{codestr},
  emphstyle=\color{codeapi},
  identifierstyle=,
  showstringspaces=false,
  breaklines=true,
  breakindent=1em,
  columns=fullflexible,
  frame=single,
  framerule=0.4pt,
  rulecolor=\color{black!45},
  numbers=left,
  numberstyle=\tiny\color{black!45},
  numbersep=5pt,
  stepnumber=1,
  xleftmargin=1.7em,
  framexleftmargin=1.7em,
  xrightmargin=0pt,
  aboveskip=2pt,
  belowskip=2pt,
}

\lstdefinestyle{vkcuda}{
  style=vkcode,
  language=C++,
  morekeywords={__global__,__device__,__shared__,__restrict__,
                __syncthreads,blockIdx,threadIdx,blockDim,gridDim},
}

\lstdefinestyle{vkpy}{
  style=vkcode,
  language=Python,
  alsoletter={.},
  emph={vk.VerifyCase,vk.input,vk.empty,vk.zeros,vk.full,vk.launch,
        vk.read,vk.write,vk.readwrite},
}

\lstdefinestyle{vkptx}{
  style=vkcode,
  language={},
  alsoletter={.},
  morecomment=[l]{//},
  morekeywords={ld.global.f32,ld.global.nc.f32,st.global.f32,add.s64,
                add.f32,sub.f32,mul.f32,fma.rn.f32,mov.u32,mov.f32,
                setp.lt.s32,bra},
}

\newcommand{\vkpanel}[1]{\par\vspace{1pt}\noindent{\footnotesize\bfseries #1}\par\vspace{2pt}}

\usepackage{xspace}

\usepackage{algorithm}
\usepackage{algpseudocode}

\usepackage{tikz}
\usetikzlibrary{arrows.meta,positioning,fit,calc}
\usepackage{pgfplots}
\pgfplotsset{compat=1.16}
\definecolor{vkNE}{HTML}{D55E00}
\definecolor{vkS}{HTML}{0072B2}
\definecolor{vkT}{HTML}{CC79A7}
\definecolor{vkTS}{HTML}{009E73}
\definecolor{vkUn}{HTML}{C9C9C9}
\definecolor{vkTo}{HTML}{444444}
\tikzset{
  vkbox/.style={draw=black!60, rounded corners=1pt, align=center,
                font=\scriptsize, inner sep=2.5pt},
  vkstage/.style={vkbox, fill=black!5},
  vkdata/.style={vkbox, fill=white},
  vkverdict/.style={vkbox, font=\scriptsize\bfseries},
  vknote/.style={align=center, font=\tiny, text=black!60},
  vkarr/.style={-{Stealth[length=4.5pt]}, black!70, thick},
}
\pgfplotsset{
  vkaxis/.style={
    tick label style={font=\scriptsize},
    label style={font=\scriptsize},
    title style={font=\scriptsize},
    legend style={font=\tiny, draw=black!25, fill=white,
                  fill opacity=0.9, text opacity=1, inner sep=1.5pt},
    axis line style={black!50},
    every tick/.append style={black!50},
    grid style={black!12},
  },
}

\newcommand{\heading}[1]{\vspace{1ex}\noindent{\bf #1}}

\newcommand{\CF}[1]{\xmakefirstuc{#1}}

\newcommand{\sys}{\textsc{Racer}\xspace}
\newcommand{\method}{symbolic equivalence checking\xspace}
\newcommand{\case}{checking instance\xspace}

\newcommand{\veq}{\texttt{Equivalent}\xspace}
\newcommand{\vneq}{\texttt{Inequivalent}\xspace}
\newcommand{\vunknown}{\texttt{Unknown}\xspace}

\newcommand{\allbugs}{82\xspace} %
\newcommand{\newbugs}{76\xspace}  %
\newcommand{\confbugs}{41\xspace}  %
\newcommand{\typefourbugs}{6\xspace}  %
\newcommand{\voltabugs}{9\xspace}  %
\newcommand{\computesanitizerbugs}{18\xspace}  %
\newcommand{\numkernels}{2{,}408\xspace}  %
\newcommand{\numrepos}{41\xspace}  %
\newcommand{\numbugrepos}{23\xspace}  %
\def\hn{\sffamily\selectfont}
\newcommand{\mpfont}{\hn\scriptsize}

\ifx\noeditingmarks\undefined
  \newcommand{\MPworker}[2]{\unskip{\color{#1}\vrule\vrule}{\marginpar{\raggedright\color{#1}\mpfont #2}}}
  \newcommand{\pgwrapper}[3]{\begingroup \color{#1} #2: #3 \endgroup}
  
  \newcommand{\pgwrapperb}[1]{\textbf{#1}}
  \newcommand{\dangerwrapper}[1]{{\color{red}#1}}
  
\else
  \newcommand{\MPworker}[2]{\unskip}
  
  \newcommand{\pgwrapperb}[1]{}
  \newcommand{\pgwrapper}[3]{}
  \newcommand{\dangerwrapper}[1]{}

\fi

\newcommand{\cheng}[1]{\pgwrapper{blue}{cheng}{#1}}
\newcommand{\tuan}[1]{\pgwrapper{red}{Tuan}{#1}}

\def\compactify{\itemsep=0in \topsep=2pt \parsep=0.00in \partopsep=0pt
  \leftmargin=2em}
\let\latexusecounter=\usecounter

{\begin{list}{\labelitemi}{\itemsep3pt \topsep3pt \parsep0.00in
          \partopsep=3pt \leftmargin1em}}%
          {\end{list}}
\newenvironment{myitemize2}%
{\begin{list}{\labelitemi}{\itemsep1pt \topsep2pt \parsep0.00in
          \partopsep=0pt \leftmargin1.2em}}%
          {\end{list}}
{\begin{list}{\labelitemi}{\itemsep1pt \topsep2pt \parsep0.00in
          \partopsep=0pt \leftmargin2em}}%
          {\end{list}}
{\begin{list}{\labelitemi}{\itemsep2pt \topsep2pt \parsep0.00in
          \partopsep=0pt \leftmargin1.2em}}%
          {\end{list}}
{\begin{list}{\labelitemi}{\itemsep3pt \topsep3pt \parsep0.00in
          \partopsep=3pt \leftmargin1.5em}}%
          {\end{list}}
{\begin{list}{\S3.3}{\itemsep3pt \topsep3pt \parsep0.00in
          \partopsep=3pt\leftmargin1em}}%
          {\end{list}}

\def\compactsortof{\itemsep=0in \topsep=2pt \parsep=0.00in \partopsep=0pt
  \leftmargin=1.2em}
\newenvironment{myenumerate2}
{\def\usecounter{\compactsortof\latexusecounter}
  \begin{enumerate}}
    {\end{enumerate}\let\usecounter=\latexusecounter}

\def\compactsqueeze{\itemsep=0pt \topsep0pt \parsep=0ex \partopsep=0pt
  \leftmargin=1.63em}

\begin{document}

\title{\sys: Hunting GPU Kernel Bugs with Symbolic Equivalence Checking}

\author{Tuan Truong Vu and Cheng Tan}
\affiliation{%
  \institution{Northeastern University}
  \city{}
  \state{}
  \country{}
}

\renewcommand{\shortauthors}{xxx}

\begin{abstract}
GPU kernels are critical to the performance of today’s AI services, yet the
same operators are rewritten repeatedly: for each GPU generation, numeric
format, serving regime, and kernel language. Every rewrite must preserve the
function computed by the implementation it replaces. Today, this obligation
depends almost entirely on \emph{differential testing}: run the new kernel and a
trusted reference on randomly sampled tensors, then compare their outputs
within a tolerance.
This approach has blind spots that additional test cases cannot eliminate. For
example, a logical bug may produce an error smaller than the tolerance
required to accommodate a correct optimized kernel.

We propose \emph{\method}, which executes a kernel and its reference on
symbolic inputs, turning each output cell into an exact algebraic expression,
and then checks whether the paired expressions agree for all inputs. Because
these expressions are exact, \method distinguishes floating-point error from
logical bugs, while a single check covers the entire input space.

We implement \sys, a PTX-level \method engine, with three techniques that make
this approach practical for production kernels:
(1) Shape Reduction to control checking cost,
(2) a lockstep CTA scheduler to handle concurrency,
and (3) idiom templates with exact numeric refutation for otherwise inconclusive results.
Across \numkernels production kernels from \numrepos collection targets, \sys found
\allbugs real bugs in \numbugrepos libraries, including \newbugs previously unknown bugs and \confbugs
confirmed so far.
\end{abstract}

\maketitle

\section{Introduction}
\label{sec:intro}

GPUs power modern AI, and GPU kernels largely determine its performance.
AI model throughput is closely tied to the throughput of its GPU kernels~\cite{wang2025systematic}.
Optimized kernels are therefore central to the performance of modern AI
systems: they directly determine inference latency and serving throughput, and
can make impractical workloads feasible. FlashAttention, for example,
substantially reduced the cost of attention and enabled efficient long-context
inference~\cite{dao2022flashattention}.
\tuan{Imo, we should mention LLMs as we say "token" or change it to "cost of serving".}
\cheng{How about now? Or you want to highlight "LLM" rather than using "AI"?}
\tuan{Yes it is good now.}

Because performance matters,
the same GPU kernel gets written over and over.
Attention alone has been rewritten for each architecture
generation as new primitives arrived,
from Ampere's asynchronous copies to Hopper's TMA and warpgroup MMA~\cite{dao2024flashattention,bikshandi2023case,chen2026tawa};
specialized per head dimension and per serving regime, prefill against decode~\cite{zhong2024distserve};
re-tuned for each numeric format~\cite{micikevicius2022fp8,frantar2025marlin,lin2023awq};
and written again in new languages such as Triton~\cite{hsu2024liger,yang2024fla,spector2025thunderkittens}.
It happens across serving platforms too:
vLLM~\cite{kwon2023efficient} and SGLang~\cite{zheng2024sglang}
each ship their own fused RMSNorm,
one written to accept any hidden size,
the other optimized for the shapes it expects to serve.

This rewriting paradigm is the norm in modern GPU kernel development: a
set of mathematical functions is implemented repeatedly, with each new
implementation striving to outperform its predecessors.

Beyond performance,
every rewrite carries a critical obligation:
it must still compute the same function as the implementation it replaces.
For today, meeting that obligation is left almost entirely
to \emph{differential testing}: run the new kernel and a trusted
implementation, usually PyTorch's, on randomly sampled input tensors, and
compare the outputs elementwise within an absolute and relative tolerance~\cite{mckeeman1998differential,paszke2019pytorch}.
This practice is widespread in production, and for good reason.
The operator set is bounded, and each operator has a precise mathematical
meaning, so a trusted reference implementation is almost always available.
Moreover, kernels are typically pure functions from input tensors to output
tensors, making side-by-side execution straightforward.

And yet, production systems still ship GPU kernels with bugs.
By studying \allbugs bugs we found in deployed libraries (\S\ref{sec:eval})
that slipped through differential testing, we identified four major obstacles:
(a) sampling cannot cover the input space:
some bugs trigger only on needle-in-a-haystack values that random sampling is unlikely to produce;
(b) not every input is an explicit argument:
many bugs depend on implicit inputs, such as uninitialized memory that happens
to contain zeros during testing but may hold arbitrary values in practice;
(c) not every output is checked:
some bugs we found corrupt memory outside the intended output region, which tests therefore never inspect;
and most notably,
(d) floating-point error can be indistinguishable from a bug:
\typefourbugs bugs we found cause only small numerical errors---smaller than the
tolerance required for correct optimized kernels.

As a real bug example,
FlagGems' InstanceNorm reconstructs the variance from the stored
$\mathrm{rstd}=1/\sqrt{\mathrm{var}+\epsilon}$ as
$1/\mathrm{rstd}^2+\epsilon$, instead of the correct
$1/\mathrm{rstd}^2-\epsilon$.
Such a bug cannot be detected by differential testing.
This is because, for typical input tensors, the output deviation caused by the
logical bug is smaller than the tolerance required to accommodate normal
floating-point error.
But this is not harmless numerical noise: the kernel computes the wrong
function, and the resulting systematic error can compound over millions of
training steps and ultimately cause training to fail.

We argue that ensuring the correctness of GPU kernels requires a more systematic and principled approach than differential testing.
To this end, we propose \emph{\method}.
\CF{\method} executes two GPU programs over symbolic inputs rather than concrete values.
As a result, each output element is represented by an algebraic expression over
free variables rather than a concrete number.
The two kernels are equivalent exactly when, for every output element, the
corresponding expressions evaluate to the same value for all possible inputs.

In \method,
each blind spot of differential testing is eliminated by design
rather than by testing harder.
First, a free variable stands for every input value at once,
so one check covers the entire space.
Second, uninitialized memory is symbolic too, so a kernel
that depends on what the previous launch left behind is checked over every
value it could have left.
Third, symbolic execution knows the address of every
store, so a write outside a declared tensor is caught where it happens.
Finally, expressions are exact rather than floating-point, removing
floating-point error from the checking problem.
Any difference between the two
kernels therefore reflects a difference in the function they compute, rather
than an artifact of rounding.

In this paper,
we introduce \sys, a \method engine for GPU kernels.
\sys takes two kernels---one under test and one reference implementation---together with a launch configuration,
and returns one of three outcomes:
\veq, \vneq{} with a concrete input that distinguishes the two kernels,
or \vunknown.

\sys executes \method at the level of PTX,
NVIDIA's virtual instruction set~\cite{ptxisa}.
PTX provides a stable interface across GPU generations and is, in fact,
the lowest-level instruction set that NVIDIA publicly documents.
As a result, many front ends, including CUDA C++ and Triton, compile to it.
Moreover, PTX exposes memory operations explicitly,
allowing the engine to model the address space that a kernel actually accesses.

Checking production GPU kernels poses three technical challenges for \sys.
First, symbolic equivalence checking is expensive at the scale of real kernels.
For example, a naive implementation of \method times out under a 600s
budget even for a simple Softmax kernel with tensor shape $4\times64$.
Second, a kernel launch is not a single piece of code running on one core, but
thousands of concurrent threads executing and interacting through
synchronization and shared memory. \sys must therefore reason not only about
each thread's arithmetic computation, but also about synchronization and
cross-thread interactions.
Finally, deciding the equivalence of arbitrary arithmetic expressions is
fundamentally difficult. In many cases, the underlying computer algebra
system %
cannot establish equivalence and returns
\vunknown. \sys must therefore reduce the frequency of such inconclusive
cases while preserving soundness.

\sys introduces three techniques to answer the three challenges:
\emph{Shape reduction} proves one representative output cell per uniformity class and transports the
proof to the rest by a verified renaming, rather than paying for every cell
(\S\ref{sec:impl:scale});
a CTA scheduler steps every thread in
lockstep between synchronization points and refuses what it cannot order,
so that no verdict rests on an interleaving the engine chose
(\S\ref{sec:impl:threads});
and idiom templates with exact numeric refutation plus a second prover for inconclusive results (\S\ref{sec:impl:decide}).

We evaluate \sys on production kernels from \numrepos GitHub repositories.
In total, \sys checks \numkernels GPU kernels.
It identifies \textbf{\allbugs real bugs}, including \textbf{\newbugs previously
unknown bugs}, of which \textbf{\confbugs have been confirmed}.
These bugs are nontrivial:
among the \allbugs bugs, an oracle differential tester could, in theory, detect only 47.
In comparison, NVIDIA's \texttt{compute-sanitizer}~\cite{nvidia2025computesanitizer}
detects \computesanitizerbugs, while Volta~\cite{volta}, the
closest existing equivalence checker, detects only \voltabugs
(\S\ref{sec:eval:compare}).

We are not the first to check GPU kernels symbolically, nor the first to
decide equivalence of tensor programs~\cite{volta,arora2025tensorright,lu2025trainverify}.
We are, to our knowledge, the first to scale \method to
bug detection in production GPU kernels: handling whole grids rather than a
single thread block, multi-launch programs, atomics, and asynchronous copies,
while achieving the performance needed to check real-world kernels.
This paper contributes:
\begin{myitemize2}
\item A study of \allbugs real bugs in production GPU kernels that identifies
four blind spots of differential testing and shows, bug by bug, why the
practice cannot close them (\S\ref{sec:motivation}, \S\ref{sec:bug}).

\item \sys, a \method engine that decides kernel equivalence at PTX level
over exact arithmetic, models GPU execution faithfully, and fails closed
rather than approximating (\S\ref{sec:design}).

\item Three techniques that make the check practical on production kernels:
shape reduction,
a lockstep CTA scheduler,
and idiom templates (\S\ref{sec:techchallenge}).

\item An evaluation on \numkernels production kernels: \allbugs real bugs across \numbugrepos
libraries, \newbugs previously unknown, 
and \confbugs confirmed at the moment.
(\S\ref{sec:eval}).
\end{myitemize2}
\section{Motivation: Kernel Correctness Today}
\label{sec:motivation}

\heading{The same operators are repeatedly rewritten.}
Unlike traditional software,
the functionality of GPU kernels for LLMs is bounded:
a few dozen operators (e.g., matrix multiply,
attention, normalization, activation) account for nearly all
of the computation a large language model performs.
Yet those same operators are written again and again: once per framework,
per hardware generation, per data type, per fusion opportunity, per
workload regime~\cite{zhong2024distserve,yu2026towards,dao2022flashattention,dao2024flashattention,bikshandi2023case,hsu2024liger,spector2025thunderkittens,yang2024fla,beck2026tiled}.
What forces the rewrites is performance.
A kernel that is correct but slow is worthless,
because the functional implementation already exists;
the only reason to write it at all is to make it faster or cheaper.
Each rewrite therefore buys speed with specialization: tiling and vectorization tuned to a shape, hand-staged
shared memory, warp-level collectives, asynchronous copies, low-precision accumulation~\cite{dao2022flashattention,dao2024flashattention,chen2026tawa,bikshandi2023case,beck2026tiled}.
Every one of those is a place to get the code wrong.

Compounding these optimizations,
the code can also go wrong because of the underlying arithmetic: tensor algebra,
whose indices, strides, reductions, and broadcast semantics
are tricky and error-prone.
An index offset by one stride, or a reduction over the wrong axis, is easy to write and difficult to spot.
For example, a buggy log-softmax kernel in FlagGems~\cite{flaggems2024}
normalizes along the tensor's last axis, even when asked to normalize
along a different one, thereby combining the wrong group of elements.
Similarly, a buggy multi-head attention kernel in attorch~\cite{attorch}
maps every head to head 0, causing all heads to compute the same output.

\heading{Differential testing is the de facto approach for correctness evaluation:}
a test samples random input tensors,
runs the newly developed kernel and a trusted reference (e.g., PyTorch's implementation)
on the same data, and compares the two outputs element-wise.
The comparison uses absolute and/or relative tolerances.
The kernel passes only if the discrepancy falls within a predefined threshold.

Differential testing fits naturally into GPU kernel development.
Because the operator set is bounded and each operator has a crisp mathematical meaning, a trusted
reference is almost always at hand, and because kernels are pure functions
from input tensors to output tensors, comparing the two is simple.
In practice, nearly all GPU kernels today ship with tests of this form.

\heading{However, differential testing misses bugs.}
Despite widespread use of differential testing,
GPU kernels continue to ship with bugs:
scanning \numkernels GPU kernels from \numrepos production-quality libraries,
we found \allbugs bugs (\S\ref{sec:eval:hunt}).
Studying those bugs, we find four reasons why differential testing is
insufficient, each a challenge that any stronger correctness check must meet.

\begin{myenumerate2}

\item \emph{Sampling cannot cover the input space.}
Testing samples insufficient input space is an old complaint, but the kernel setting amplifies it.
A single argument is a tensor of thousands
of elements, and an element is not drawn from a handful of categories but
from a floating-point domain of billions of bit patterns.
The combination of the two makes good coverage infeasible.
Meanwhile, we do find bugs that manifest only for specific input tensor values,
for example,
attorch's hardshrink keeps an input equal to exactly $\pm\lambda$, where
PyTorch's keep-set is strict ($|x|>\lambda$); the two differ only where an
input element equals the threshold $\lambda$ exactly (default $0.5$).

\item \emph{Not every input is an argument.}
Further, a kernel's result depends on state that never appears in its signature.
Some kernels depend on allocated but uninitialized memory, which often happens
to contain zeros during testing but is not guaranteed to be zero in production.
For example, a reduction kernel in
Triton-distributed~\cite{zheng2025tritondistributed} has each CTA atomically
add its partial sum into an output cell that is only allocated but not
initialized as zero, so the result is the sum plus whatever the cell held
before the launch.

\item \emph{Not every output is checked.}
A test compares the tensors the kernel declares; the kernel writes wherever
its computed addresses point. A miscomputed index or a missing bounds guard
on a ragged final tile scribbles outside the intended region, into a
neighboring tensor or into a buffer another kernel owns, while the declared
output still matches the reference element for element.
For example,
fla's squared-ReLU~\cite{yang2024fla} backward writes past the tensor end when the length is
not a multiple of block size.

\item \emph{Floating-point error is indistinguishable from a bug.}
Most importantly,
numeric thresholds cannot distinguish an incorrect function from a correct one evaluated under a different floating-point order, leading to errors in both directions.
A correct kernel may exceed the tolerance due to reduction reordering; more seriously, an incorrect kernel may remain within the tolerance on sampled inputs.
Loosening the tolerance suppresses false alarms but can also mask real bugs.
We have found \typefourbugs cases in this category, where incorrect logic is hidden by the floating-point error threshold.

\end{myenumerate2}

\noindent
These four points lead to one conclusion:
a test compares numbers, but correctness
is a statement about functions.
What the setting calls for is a check that
ranges over all inputs rather than sampled ones, that accounts for the
memory a kernel actually touches rather than the tensors it declares, and
that decides whether two implementations compute the same value rather than
producing nearby ones.

\section{Equivalence Checking for GPU Kernels}
\label{sec:background}

\subsection{Beyond Testing: Toward Principled Checking}
\label{sec:background:idea}

\cheng{for me, the below two paragarphs are not too different from the two in
intro; people may feel repetitive}

We believe testing cannot address the four challenges described in \S\ref{sec:motivation}.
Our answer is \emph{\method}.

\heading{\CF{\method}.}
This method executes a kernel over \emph{symbolic} inputs, so each instruction
builds an algebraic expression instead of computing a number, and every
output tensor holds one expression per cell rather than one value.
Running the new kernel and its reference this way yields, for every
output cell, a pair of expressions over the same input variables.
The two kernels are \emph{equivalent} when both expressions result in the same value
for all possible inputs.
Nothing is sampled and compared numerically.

All four challenges dissolve at this level.
\emph{Input coverage}:
a free variable stands for every value at once, so a single check covers the
entire input space.
\emph{Implicit inputs}: the cells of an
uninitialized buffer are free variables too, so the check ranges over
whatever the running context may have left in that memory.
\emph{Unchecked outputs}: the execution records the
address of every store, and each expression records which inputs reached the
cell, so a write outside a declared tensor is caught where it happens instead
of surfacing later as someone else's wrong number.
\emph{Floating-point error}: expressions are exact rather than
floating point, so reordering a reduction changes an expression's shape but
not its value.
There is no tolerance parameter to tune; any difference between two expressions
corresponds to a difference in the computed function, not to floating-point rounding.

\heading{Prior equivalence checkers.}
The idea of proving two implementations equivalent is not new, and
\S\ref{sec:relwork} returns to it.
For example,
TrainVerify~\cite{lu2025trainverify} checks a distributed training
program against its single-device specification;
TensorRight~\cite{arora2025tensorright}
applies equivalence reasoning to tensor-graph rewrites, certifying that a
rewritten graph computes what the original did;
Volta~\cite{volta}, the closest to us,
brings equivalence checking to GPU kernels themselves.
\sys differs from prior work in its focus on finding bugs:
of the \allbugs bugs \sys finds, Volta detects only \voltabugs.
We explain this gap in \S\ref{sec:eval:compare}.

\subsection{Background: GPU Execution and PTX}
\label{sec:background:execmodel}

GPU programs have distinctive characteristics that shape how \method must be
built. This subsection reviews the basics of GPU execution, then PTX~\cite{ptxisa},
the instruction set at which \sys operates.

\heading{GPU execution.}
A GPU is a massively parallel processor that runs tens of thousands of
threads at once. The program those threads run is a \emph{kernel}: a single
function, written from the perspective of one thread, that every thread
executes over its own slice of the data. To run a kernel, the CPU host
\emph{launches} a grid of cooperative thread arrays (CTAs), also called
thread blocks; the grid and block dimensions, chosen at launch time
together with the kernel's arguments, form the \emph{launch
configuration}. Every thread runs the same kernel body and is distinguished
only by its block and thread indices: the single-instruction,
multiple-thread (SIMT) model. Threads advance in groups of 32, called
\emph{warps}, whose members can exchange values directly.

A kernel reads and writes three memory spaces of different scope and
lifetime: registers and local memory, private to a thread; shared memory, a
per-CTA scratchpad that dies with the CTA; and global memory, visible to
every thread in the launch and persistent across launches. Threads within a
CTA synchronize with a barrier (\texttt{bar.sync}), but threads in different
CTAs cannot synchronize within an ordinary launch, so a computation needing
a grid-wide handoff is split into several launches, each reading through
global memory what the previous one wrote.
Split-K matrix multiplication is
the standard example: one launch has each CTA reduce its own slice of K into
a partial-product buffer, and a second launch sums those partials.
Both facts matter later: the engine must reproduce this ordering discipline
faithfully %
(\S\ref{sec:impl:threads}).

\heading{PTX.}
\label{sec:background:ptx}
PTX is NVIDIA's stable virtual instruction set. Kernels written in different
front ends, such as CUDA C++~\cite{cudaguide}, Triton~\cite{tillet2019triton}, Numba~\cite{lam2015numba}, TVM~\cite{chen2018tvm}, and CuPy~\cite{cupy_learningsys2017},
all lower to PTX before reaching the GPU, which makes it the one level at
which kernels from different front ends can be compared instruction for
instruction.
For example,
Figure~\ref{fig:lowering}
shows one add-and-narrow computation in CUDA C++
and Triton, both of which are compiled to
the same PTX instructions.

\begin{figure}[t]
\begin{lstlisting}[language=C]
// CUDA C++ (nvcc)
h[i] = __float2half_rn(x[i] + y[i]);

# Triton (MLIR/LLVM)
tl.store(H + i, (tl.load(X + i)
                 + tl.load(Y + i)).to(tl.float16))

// PTX (both front ends compile to the same PTX)
add.rn.f32      %
cvt.rn.f16.f32  %
st.global.b16   [%
\end{lstlisting}
\caption{Add two fp32 values and narrow to fp16, in CUDA C++, in Triton, and
the PTX both compile to.}
\Description{CUDA and Triton snippets for the same computation and the
three PTX instructions both lower to.}
\label{fig:lowering}
\end{figure}

\subsection{Challenges in Applying Symbolic Execution}
\label{sec:background:symbolic}

\begin{figure}[th]
\begin{lstlisting}
A:  t = x + y;  out = t * t
B:  out = x*x + 2*x*y + y*y
\end{lstlisting}
\vspace{-4pt}
{\scriptsize
\[
e_A = (x+y)^2, \qquad e_B = x^2+2xy+y^2,
\]
\[
\mathrm{simplify}(e_A - e_B) \to 0
\;\;\Rightarrow\;\; \forall x,y, \, A \equiv B
\]}
\caption{Symbolic execution and CAS equality: each program yields an
expression in the free inputs, and their difference simplifying to zero
proves their equivalence.}
\label{fig:symexec}
\end{figure}
 
\heading{Symbolic execution and computer algebra.}
\emph{Symbolic execution} runs a program over free variables instead of
concrete data: each arithmetic instruction builds an algebraic expression in
those variables rather than computing a number, and each branch on a
symbolic condition is resolved under the constraint that selects its path.
A \emph{computer algebra system} (CAS)~\cite{Sympy,Mathematica,Maple,SageMath,Maxima}
then manipulates those expressions exactly,
using unbounded-precision rational arithmetic
and algebraic rewriting instead of evaluating them as floating-point approximations.
We use SymPy~\cite{Sympy}, a CAS implemented as a Python library.
To check whether two expressions are equivalent,
we simplify their difference and check whether it reduces to zero.
Figure~\ref{fig:symexec} shows an example.

\heading{Symbolic executing PTX code.}
Applying symbolic execution on PTX code yields a recipe for \method.
Bind every cell of every input tensor to a free variable instead of a number. Execute the
kernel's PTX one instruction at a time, interpreting each by the mathematical
operation it denotes in the CAS: a floating-point instruction becomes the
corresponding operation over the reals.
When the launch ends, every output cell holds an
exact expression in the input variables, and no rounding has been lost to an
approximation along the way. Checking two kernels is then subtraction: run
both over the \emph{same} symbolic inputs, form the difference of their
expressions cell by cell, and ask the CAS whether every difference is zero.
A logical bug is found if the two kernels disagree.

\heading{Challenges.}
Naively applying a CAS to PTX instructions, however, does not work.
Three challenges arise in practice.

\begin{myitemize2}

\item \emph{Scalability.}
\label{sec:challenges:scale}
A GPU kernel launch can contain thousands of threads, each performing thousands
of arithmetic operations. Executing all threads and proving each output
cell independently scales poorly with both thread count and expression
size: an output at the end of a long reduction carries the full
expression chain that produced it, which the CAS must repeatedly simplify.

\item \emph{Concurrency.}
GPU kernels are highly concurrent, with many threads cooperating to complete a computation.
A symbolic engine must therefore model concurrency primitives such as
barriers,
warp collectives,
asynchronous copies,
atomic instructions,
and tensor-core instructions.
Moreover, the result may depend on thread interleavings, so the engine must
detect and reason about data races. Each construct must either be
modeled faithfully or rejected outright (\S\ref{sec:impl:threads}).

\item \emph{Inconclusive results.}
Deciding whether two symbolic expressions are equal is a hard problem.
For expressions involving a sufficiently rich set of functions, modern CASs may fail to reach a conclusion or may time out.
In practice, this happens often.
We therefore need to minimize the chance that the engine returns \vunknown (\S\ref{sec:impl:decide}).

\end{myitemize2}
\section{\sys: a Symbolic Equivalence Checker}
\label{sec:design}

\begin{figure*}[t]
\centering
\footnotesize
\begin{minipage}[t]{0.485\textwidth}
\vkpanel{(a) The two kernels (CUDA C++)}
\begin{lstlisting}[style=vkcuda]
// reference: GPT-J rotates the pairs (2i, 2i+1)
__global__ void rope_gptj(float* out, const float* x,
    const float* cosv, const float* sinv, int D) {
  int t = blockIdx.x;
  for (int i = 0; i < D / 2; i++) {
    float a = x[t*D + 2*i], b = x[t*D + 2*i+1];
    float c = cosv[i], s = sinv[i];
    out[t*D + 2*i]   = a*c - b*s;
    out[t*D + 2*i+1] = a*s + b*c;
  } }

// shipped for GPT-J heads (#941): the NeoX pairing
__global__ void rope_neox(...) { // same prologue
  for (int i = 0; i < D / 2; i++) {
    float a = x[t*D + i], b = x[t*D + i + D/2];
    float c = cosv[i], s = sinv[i];
    out[t*D + i]       = a*c - b*s;
    out[t*D + i + D/2] = a*s + b*c;
  } }
\end{lstlisting}

\vkpanel{(c) The PTX loop bodies}
\begin{lstlisting}[style=vkptx]
// rope_gptj: one pointer, stepping 8B per iteration
$L__BB0_2:
  ld.global.f32 %
  ld.global.f32 %
// rope_neox: two pointers (D/2)*4B apart, each 4B
$L__BB1_2:
  ld.global.f32 %
  ld.global.f32 %
\end{lstlisting}
\end{minipage}\hfill
\begin{minipage}[t]{0.485\textwidth}
\vkpanel{(b) The \case (Python)}
\begin{lstlisting}[style=vkpy]
import vk

class RopeGptjNeox(vk.VerifyCase):
    T, D = 2, 4          # tokens, head dim

    def setup(self):
        self.x    = vk.input("x", (self.T, self.D))
        self.cosv = vk.input("cosv", (self.D // 2,))
        self.sinv = vk.input("sinv", (self.D // 2,))
        self.out  = vk.empty("out", (self.T, self.D))

    def _run(self, kernel):
        vk.launch(
            source="rope.cu", kernel=kernel,
            grid=(self.T, 1, 1),
            block=(1, 1, 1),
            args=[vk.write(self.out),
                  vk.read(self.x),
                  vk.read(self.cosv),
                  vk.read(self.sinv),
                  self.D])
        return {"out": self.out}

    def program_a(self):     # the shipped kernel
        return self._run("rope_neox")

    def program_b(self):     # the reference
        return self._run("rope_gptj")
\end{lstlisting}
\end{minipage}
\caption{The running example, vLLM \#941.
\textbf{(a)} two CUDA rotary-embedding kernels that differ only in which
elements they pair;
\textbf{(b)} the \case declaring both over the same symbolic inputs at one
launch configuration;
\textbf{(c)} the PTX loop bodies the two kernels lower to: same arithmetic,
different addresses.}
\Description{Three code panels. Panel (a) gives two CUDA
rotary-embedding kernels that differ only in which elements they pair,
panel (b) the Python checking instance declaring both over the same
symbolic input tensors at one launch configuration, and panel (c) the
PTX loop body each kernel lowers to.}
\label{fig:case}
\end{figure*}

\noindent
\sys answers the question:
\emph{do the two given GPU kernels compute the same function?}
A user poses this question as a \emph{\case} containing two GPU programs---the
second serving as the reference implementation, as in differential testing---the
tensors they read and write, and their launch configurations.
\sys runs the \case.
When done,
\sys returns one of three verdicts: \veq,
\vneq with a concrete input that separates the two,
or \vunknown.

It takes four steps for \sys to produce the verdict.
\sys \emph{(1)} reads the \case,
\emph{(2)} lowers both programs to PTX whatever front end they were written in,
\emph{(3)} executes every thread of every CTA of both programs
symbolically, leaving one algebraic expression in each output cell,
and \emph{(4)} decides, cell by cell, whether
the two expressions denote the same value.
The rest of this section walks the four steps on a running example in
Figure~\ref{fig:case}, a bug from vLLM (\#941).
\label{sec:design:interface}
\heading{Step 1: posing the question as a \case.}
In \sys,
a \case is a Python class with three methods.
The first, \texttt{setup}, declares the tensors both sides share;
the other two, \texttt{program\_a} and \texttt{program\_b}, each declare one program as a
sequence of launches.
Tensor declarations specify shapes, data types, and initial contents:
\texttt{vk.input()} introduces a free variable for each cell,
\texttt{vk.empty()} leaves cells uninitialized, and
\texttt{vk.zeros()} or \texttt{vk.full()} initializes cells
to a specified constant.
Each program records a sequence of kernel launches through
\texttt{launch} call, specifying the kernel, its arguments,
and its launch configuration.
Tensor arguments are wrapped in \texttt{vk.read},
\texttt{vk.write}, or \texttt{vk.readwrite} to declare
how each launch accesses them.
Successive launches exchange data by accessing the same tensors.
Each program method returns a dictionary naming the output tensors
whose corresponding cells \sys should compare.

\cheng{Tuan, check if the above paragraph is correct.}
\tuan{(i) There are also vk.zeros, vk.full. 
(ii) Minor in the last clause: the launch does not return dictionary.}
\cheng{Okay. Tuan, can directly modify the above paragraph? (don't have to use your micro)}
\tuan{edited}

In Figure~\ref{fig:case}(b),
\texttt{setup} declares the tensors:
a $2{\times}4$ activation \texttt{x} and the rotation tables
\texttt{cosv} and \texttt{sinv} as inputs (free variable)
and the output \texttt{out} as \texttt{empty}. Each program is a single
launch of one of the two kernels over the same arguments: one CTA per
token, one thread per CTA, the head dimension $D{=}4$ passed as a scalar,
and \texttt{out} wrapped as \texttt{vk.write} where the inputs are
\texttt{vk.read}.
The \case poses the question: at this
configuration, do these two kernels compute the same function of
\texttt{x}, \texttt{cosv}, and \texttt{sinv}?

\label{sec:design:frontends}
\heading{Step 2: lowering GPU programs to PTX.}
Each launched kernel is lowered to PTX before analysis, making all subsequent stages front-end agnostic (\S\ref{sec:background:ptx}).
The current \sys implementation supports three input forms: CUDA C++, Triton, and PTX.
In principle, \sys can support any GPU programming language whose toolchain compiles to PTX.
\sys only requires a language-specific adapter.

In our example, both programs are written in CUDA and lower to different PTX
instructions: Figure~\ref{fig:case}(c) excerpts the loop body.
The bug lives in the address arithmetic.
\texttt{rope\_gptj} loads adjacent elements from
\texttt{[\%rd40+-4]} and \texttt{[\%rd40]}, advancing
\texttt{\%rd40} by 8 bytes per iteration.
\texttt{rope\_neox} instead loads from \texttt{[\%rd45]}
and \texttt{[\%rd39]}, where \texttt{\%rd39} is computed as
\texttt{\%rd45} plus $(D/2)\cdot4$ bytes.
Advancing \texttt{\%rd45} by 4 bytes per iteration therefore
pairs elements from the two halves of the head.
\tuan{Fix the wrong registers name in ~\ref{fig:case}(c) and updated the text.}

\label{sec:design:engine}
\heading{Step 3: executing PTX symbolically.}
\sys implements a symbolic execution engine that interprets each PTX
instruction into a CAS representation.
Across threads, \sys uses a lockstep scheduler to coordinate execution within
each thread block.
For example, consider the first iteration ($i=0$) of
\texttt{rope\_}\allowbreak\texttt{neox} for token $t$, with $D=4$.
Address calculations resolve to concrete input and output cells,
while the values loaded from inputs remain symbolic.
The following instruction sequence illustrates how \sys constructs
the expression for \texttt{out[$t$,0]}.
Each row shows an instruction and its effect on symbolic state,
with floating-point arithmetic interpreted over the reals:
\begin{center}\footnotesize
\begin{tabular}{@{}l@{\quad}l@{}}
\textit{PTX instruction} & \textit{Symbolic state update}\\
\texttt{ld.global.f32 \%f37,[\%rd47]} & $\texttt{\%f37} \gets c_0$\\
\texttt{ld.global.f32 \%f38,[\%rd45]} & $\texttt{\%f38} \gets x_{t,0}$\\
\texttt{mul.f32 \%f39,\%f38,\%f37} & $\texttt{\%f39} \gets c_0\,x_{t,0}$\\
\texttt{ld.global.f32 \%f40,[\%rd48]} & $\texttt{\%f40} \gets s_0$\\
\texttt{ld.global.f32 \%f41,[\%rd39]} & $\texttt{\%f41} \gets x_{t,2}$\\
\texttt{mul.f32 \%f42,\%f41,\%f40} & $\texttt{\%f42} \gets s_0\,x_{t,2}$\\
\texttt{sub.f32 \%f43,\%f39,\%f42} & $\texttt{\%f43} \gets c_0x_{t,0}-s_0x_{t,2}$\\
\texttt{st.global.f32 [\%rd46],\%f43} & $\texttt{out}[t,0] \gets c_0x_{t,0}-s_0x_{t,2}$\\
\end{tabular}
\end{center}
Each arithmetic instruction builds an expression from those already
held in its source registers; the store writes the resulting
expression to the addressed output cell.
In \texttt{rope\_gptj}, the corresponding second activation load
reads $x_{t,1}$ instead of $x_{t,2}$, producing
$c_0x_{t,0}-s_0x_{t,1}$.

\label{sec:design:check}
\heading{Step 4: checking output cells.}
When both programs have run, their declared output tensors are paired by
name, giving for every output cell a pair of expressions to compare.
The checker forms the difference of each pair and hands it to the underlying CAS (\S\ref{sec:background:symbolic}).
A \case is \veq only if every cell of
every output tensor simplified to zero; a single refuted cell fails the case.
\cheng{Tuan, why it called prescreen? isn't this happen after the CAS's simplification?}\tuan{Initially, I have it before the simplification, and its name is prescreen in the code.
When I wrote the design, I think putting it after the simplification will make the pipeline more natural, so I switched it.}
If the CAS leaves a cell unresolved, \sys applies exact numeric refutation,
which substitutes a concrete rational assignment for the free variables and evaluates expressions exactly:
a non-zero result is an \vneq witness,
and its absence leaves the cell \vunknown,
since exact evaluation at one point can
refute but never prove.
\tuan{May need to add more detail about prescreening.}

In the running example, for cells at position $[t,0]$,
\[
    e_{a}[t,0] = c_0\,x_{t,0} - s_0\,x_{t,2},
\quad
\overset{?}{=}
\quad
e_b[t,0] = c_0\,x_{t,0} - s_0\,x_{t,1}.
\]
Their difference is
$e_a-e_b=s_0\,(x_{t,1}-x_{t,2})$.
Exact numeric refutation establishes that this expression is not
identically zero: substituting $s_0{=}1$, $x_{t,1}{=}1$, and
$x_{t,2}{=}0$ yields $e_a-e_b=1$.
This assignment is a concrete witness, so \sys returns \vneq.

\cheng{for me, a type 4 bug in the latex; think of where to put it}
\section{Technical Challenges}
\label{sec:impl}
\label{sec:techchallenge}

The design in \S\ref{sec:design} must address the three challenges
introduced in \S\ref{sec:background:symbolic}: scalability, concurrency,
and inconclusive results. This section explains how \sys addresses them
with shape reduction, a lockstep CTA scheduler, and idiom templates
with exact numeric refutation and a second prover based on Z3.

\subsection{Scalability}
\label{sec:impl:scale}

To prove equality at output cell $c$, the CAS attempts to simplify
$e^a_c - e^b_c$ to zero using algebraic transformations such as expanding
products, cancelling factors, and combining fractions. This process can
be costly for large expressions. For an output tensor with thousands of cells,
deciding each cell independently repeats this work even when the
expression pairs differ only in the input variables they reference.
\emph{Shape reduction} reduces the deciding cost by reusing proofs across
output cells with the same expression structure.

\heading{Reusing a representative's proof.}
GPU kernels commonly apply the same function to different slices of an
input tensor. Their output expressions then differ only in the input
variables they reference. Shape reduction groups output cells into uniformity classes: their expression pairs match under a shared renaming of input variables.

Let $e^a_r$ and $e^b_r$ be the expressions produced by the two programs
at a representative cell $r$, and suppose the deciding procedure has proved
$e^a_r = e^b_r$. Another cell $c$ can reuse this proof if a renaming
$\tau$ satisfies
\[
  e^a_c = \tau(e^a_r)
  \quad\text{and}\quad
  e^b_c = \tau(e^b_r),
\]
hence $e^a_c = e^b_c$.

For each output tensor, \sys processes cells in sequence and tries to
reuse a proved representative. It validates each candidate renaming by
comparing both expression pairs structurally, allowing reordering of
commutative arguments without invoking CAS simplification.

\begin{figure}[t]
\centering
\begin{tikzpicture}[x=1mm,y=1mm,font=\scriptsize]
  \node[anchor=west,inner sep=0pt] at (0,24) {(a) $C=AB$};
  \node at (3.45,20) {$A$};
  \node at (16.6,20) {$B$};
  \node at (30.6,20) {$C$};
  \foreach \r in {0,...,3} \foreach \c in {0,...,2}
    \draw[black!45] (\c*2.3,15-\r*2.3) rectangle ++(2.3,2.3);
  \foreach \c in {0,...,2} {
    \draw[draw=black!45,fill=vkS!35] (\c*2.3,15) rectangle ++(2.3,2.3);
    \draw[draw=black!45,fill=vkS!12] (\c*2.3,10.4) rectangle ++(2.3,2.3);
  }
  \foreach \r in {0,...,2} \foreach \c in {0,...,3}
    \draw[black!45] (12+\c*2.3,12.7-\r*2.3) rectangle ++(2.3,2.3);
  \foreach \r in {0,...,2} {
    \draw[draw=black!45,fill=vkTS!35] (12,12.7-\r*2.3) rectangle ++(2.3,2.3);
    \draw[draw=black!45,fill=vkTS!12] (16.6,12.7-\r*2.3) rectangle ++(2.3,2.3);
  }
  \foreach \r in {0,...,3} \foreach \c in {0,...,3}
    \draw[black!45] (26+\c*2.3,15-\r*2.3) rectangle ++(2.3,2.3);
  \draw[fill=black!70] (26,15) rectangle ++(2.3,2.3);
  \draw[very thick,vkNE] (30.6,10.4) rectangle ++(2.3,2.3);
  \node[anchor=west,inner sep=0pt] at (0,3) {$A[0,k]\mapsto A[i,k]$};
  \node[anchor=west,inner sep=0pt] at (0,-2) {$B[k,0]\mapsto B[k,j]$};

  \begin{scope}[xshift=43mm]
    \node[anchor=west,inner sep=0pt] at (0,24) {(b) $Y[i,j] = e^{X[i,j]} \big/ \sum_k e^{X[i,k]}$};
    \node at (4.6,20) {$X$};
    \node at (28.6,20) {$Y$};
    \foreach \r in {0,...,3} \foreach \c in {0,...,3} {
      \draw[black!45] (\c*2.3,15-\r*2.3) rectangle ++(2.3,2.3);
      \draw[black!45] (24+\c*2.3,15-\r*2.3) rectangle ++(2.3,2.3);
    }
    \foreach \c in {0,...,3} {
      \draw[draw=black!45,fill=vkS!35] (\c*2.3,15) rectangle ++(2.3,2.3);
      \draw[draw=black!45,fill=vkS!12] (\c*2.3,10.4) rectangle ++(2.3,2.3);
    }
    \draw[draw=black!45,fill=vkTS!35] (0,10.4) rectangle ++(2.3,2.3);
    \draw[draw=black!45,fill=vkTS!35] (4.6,10.4) rectangle ++(2.3,2.3);
    \draw[<->,thick,vkTS!70!black] (1.15,7.4) to[bend right=35] (5.75,7.4);
    \draw[fill=black!70] (24,15) rectangle ++(2.3,2.3);
    \draw[very thick,vkNE] (28.6,10.4) rectangle ++(2.3,2.3);
    \node[anchor=west,inner sep=0pt] at (0,3) {$X[0,k]\mapsto X[i,k]$};
    \node[anchor=west,inner sep=0pt] at (0,-2) {then $X[i,0]\leftrightarrow X[i,j]$};
  \end{scope}
\end{tikzpicture}
\caption{Candidate renamings from representative (black) to target
(outlined) cells.}
\label{fig:shape-reduction}
\end{figure}
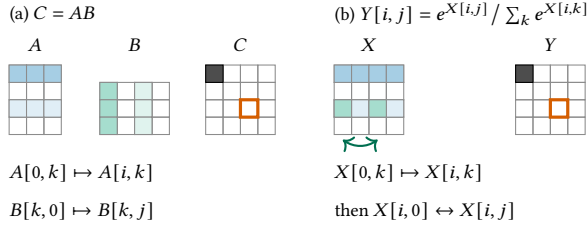
 
\heading{Finding candidate renamings.}
Candidates come from input \emph{footprints}: the indices in each input
tensor referenced by either program's expression at a cell. \sys first
proposes a translation that maps each representative footprint onto the
target footprint. When the output position identifies an element within
an input footprint, it also proposes exchanging the translated
representative position with the target position.
Figure~\ref{fig:shape-reduction} illustrates both patterns: matrix
multiplication translates a row of $A$ and a column of $B$; row softmax
translates a row of $X$ and exchanges two entries to select the target
numerator while preserving the denominator's sum. Footprints only
generate candidates; proof reuse requires the structural validation above.
Algorithm~\ref{alg:shape} in Appendix~\ref{sec:appendix:algorithms}
gives the candidate-generation procedure.

\subsection{Concurrency}
\label{sec:impl:threads}

Threads can compute correct expressions individually and still produce
an incorrect result if one reads a value before another writes it.
To address the concurrency challenge in
\S\ref{sec:background:symbolic}, \sys models both what each thread
computes and when threads can exchange values.

\heading{Coordinating threads.}
\sys executes every thread symbolically. Within each CTA, its scheduler
runs a thread until it exits or reaches a synchronization point.
At a barrier, the thread waits until all required participants arrive;
only then can they continue. The scheduler also coordinates the
participants of supported warp and tensor-core collectives. If required
threads cannot meet at the same synchronization point, execution is
refused. Thus, a thread cannot pass a barrier simply because the engine
happened to execute it first.
Algorithm~\ref{alg:scheduler} in Appendix~\ref{sec:appendix:algorithms}
gives the scheduler pseudocode.

\heading{Ordering memory accesses.}
Scheduling alone cannot detect a missing barrier: running a writer
before a reader could hide the error. \sys therefore records reads and
writes in \emph{epochs}, the intervals between CTA-wide barriers, and
reports conflicting accesses that lack the required ordering. A CTA
barrier orders accesses within that CTA; conflicts across CTAs are
tracked separately, and global memory is retained across launches.
Asynchronous copies require another ordering rule: issuing a copy does
not make its data available. \sys exposes the copied data only at the
modeled completion point---the corresponding wait for \texttt{cp.async},
or completion of the associated barrier for TMA---and refuses accesses
that would use it earlier.

\heading{Combining atomic updates.}
Multiple threads may atomically update the same cell in different
orders. \sys collects these updates and combines them at the next CTA
barrier for shared memory, or at launch completion for global memory.
This is valid only when the final value is independent of update order
under the arithmetic model, as with addition over exact reals
(\S\ref{sec:background:symbolic}) and supported integer add, min, max,
and bitwise operations. The engine refuses uses of an atomic's returned
old value, mixtures of atomic operations on one cell, and ordinary
accesses that conflict with pending updates.

\subsection{Inconclusive Results}
\label{sec:impl:decide}

The CAS may fail to prove equality within its time budget even when two
expressions denote the same function (\S\ref{sec:background:symbolic}).
\sys uses three techniques to reduce these inconclusive outcomes.

\heading{Idiom templates.}
Compilation can turn a familiar function, such as $\log(x)$, into a
long sequence of bit operations and polynomial arithmetic that obscures
its meaning to the CAS. \sys recognizes supported routines using
templates that match their coefficients, operations, guards, and
register dependencies. It replaces each matched sequence with the
corresponding mathematical operation under the real-arithmetic model
of \S\ref{sec:background:symbolic}.
For example, an inlined libdevice \texttt{\_\_nv\_logf} contains the
following PTX instructions:
\begin{lstlisting}[style=vkptx,numbers=none]
mov.b32 %
add.s32 %
and.b32 %
// ...
mov.f32 %
fma.rn.ftz.f32 %
// ...
setp.lt.u32 %
// ...
selp.f32 %
\end{lstlisting}
Matching the full
routine lets \sys represent its result in \texttt{\%f37} as $\log(x)$
for a positive input $x$, rather than carrying this instruction sequence
into the output comparison.

\heading{Exact numeric refutation.}
If the CAS leaves a cell unresolved, \sys evaluates both expressions
at the same concrete input: rational values for free real inputs and
integers for index inputs, satisfying the \case's constraints and the
expressions' domains. It computes their difference using exact fractions
for rational arithmetic. For functions such as square roots, exponentials,
and logarithms, it uses interval evaluation with outward rounding to
bound the true difference. A nonzero exact difference, or an interval
excluding zero, yields an \vneq witness; otherwise, the cell remains
unresolved.

\heading{A second prover.}
When both the CAS and numeric refutation leave a cell unresolved, \sys
uses Z3~\cite{demoura2008z3} to search for an input where its expressions
differ. This can target boundaries that random inputs rarely reach:
$x\geq\lambda$ and $x>\lambda$, for example, differ only at
$x=\lambda$. For supported expressions, the query encodes both outputs
and the \case's input constraints. A proposed counterexample yields
\vneq only after the numeric procedure above confirms a nonzero
difference in the original expressions. An unsatisfiable query proves
equality only if the encoding preserves the expressions' arithmetic
semantics and input domain. Like a CAS proof, this proof can be reused
by shape reduction after validating the shared renaming. Unsupported
translations, timeouts, inconclusive solver results, and unvalidated
candidates leave the cell unresolved.

A concrete refutation yields \vneq. Otherwise, the \case is \veq
only when every output cell has been proved equal, directly or by proof
reuse; any remaining unresolved cell makes it \vunknown.
For an \vunknown case, we manually inspect the kernel code alongside
the unresolved symbolic expressions. This helps us identify bugs or
limitations of the deciding procedure. %
\section{Bug Study}
\label{sec:bug}

Our study of \allbugs bugs across \numbugrepos libraries, including \newbugs
previously unknown bugs and six rediscoveries, reveals how real GPU
kernel defects escape testing. We assign each bug to one category based
on the primary reason a test can overlook it, illustrating the four
limitations in \S\ref{sec:motivation} before describing the other findings.

\heading{Sampled input.}
Some bugs manifest only at a special input value that random tests rarely
produce, even when they exercise the faulty code.

For example, hardshrink should return zero when $|x|\leq\lambda$
and leave the input unchanged otherwise. The attorch implementation
instead keeps values equal to the threshold: with $\lambda=0.5$,
it returns $0.5$ for input $0.5$ instead of zero. Inputs on either
side give correct results, so exposing the bug requires hitting the
boundary. The seven findings in this category require values rarely
produced or omitted by the studied generators, including special labels
that mark elements to ignore.

\heading{Implicit input.}
A kernel's result can depend on memory state or execution conditions
that tests leave unchanged while varying tensor values.

For example, a Triton-distributed reduction allocates an output cell
with \texttt{torch.empty} and atomically adds each block's partial sum
without first setting the cell to zero. The cell may retain data from
an earlier use: if it initially contains $c$, the result is $s+c$
instead of the intended sum $s$. Atomic addition prevents lost updates
but does not initialize the cell. \sys reports an uninitialized-memory
diagnostic, which disappears after zero-initialization.
The 11 findings in this category also involve tensor layouts and
parameter settings left unvaried by the studied tests, even when these
are explicit arguments.

\heading{Unchecked output.}
A kernel can compute the output tensor correctly yet write past its
allocated buffer. A test comparing only that tensor does not inspect
the extra writes.

For example, fla's squared-ReLU backward kernel handles 4{,}096
elements per block and should limit the final block's writes to valid
elements. Instead, it writes all 4{,}096 results even when the output
has only 64 elements. Those 64 values match the reference, but the
remaining 4{,}032 writes exceed the output allocation and overwrite
memory beyond it. This category contains six out-of-bounds write bugs
across five libraries.

\heading{Below tolerance.}
A wrong formula can produce values close enough to the reference to
pass a test: tolerances intended to allow floating-point rounding can
also hide programming errors.

For example, FlagGems' InstanceNorm recovers variance from a stored
value $\mathrm{rstd}=1/\sqrt{\mathrm{var}+\epsilon}$.
Squaring and inverting gives $\mathrm{var}+\epsilon$, so recovery
requires subtracting $\epsilon$. The kernel adds it instead,
overestimating the variance by $2\epsilon$ before updating the running
statistics. For small $\epsilon$, the resulting error can fall within
the test's tolerance. The six findings in this category are masked by
the tolerances used for the studied inputs.

\heading{Other.}
Other findings involve gaps in test coverage, invalid memory accesses,
and synchronization errors. For example, the rotary-embedding kernel
in Figure~\ref{fig:case} pairs the wrong input elements.
Of these 52 bugs, 40 can be exposed by comparison
with an independent reference at a triggering configuration; existing
tests miss relevant paths, share faulty logic, or omit a comparison.
Seven read outside an allocated buffer, e.g. FlagGems' bucketize fetches elements past the input's end.
Four are data races: threads access the same memory, with
at least one writing, without synchronization to order the accesses.
Their effects depend on thread timing, so a passing run does not
establish correct synchronization.
One further bug modifies padding within the allocation without changing the operator's semantic output.
\section{Evaluation}
\label{sec:eval}

We evaluate \sys's bug findings, detection by other methods,
equivalence coverage, and end-to-end checking cost.

\heading{Experimental setup.}
We conduct the experiments on a server with two AMD EPYC 9575F CPUs
(128 cores), Python 3.11.16, and SymPy 1.13.1. An NVIDIA RTX PRO 6000
captures launches, replays witnesses, and runs
the concrete baselines. The server runs Rocky Linux 9.6 with CUDA 13.1. 
\sys compiles CUDA sources with nvcc 12.4 and
Triton kernels with Triton 3.2.0, both emitting PTX ISA 8.4.

\subsection{Finding Bugs in Production Kernels}
\label{sec:eval:hunt}

We examine production kernels from \numrepos collection targets with \sys
and identify \allbugs distinct bugs across 23 libraries:
\newbugs previously unknown bugs and six rediscoveries.
\S\ref{sec:bug} analyzes the defects; here we describe
how they were found and counted.

\heading{Pipeline.}
We first harvest kernel entry points from open-source implementations
and capture the kernel together with the grid, block,
and shared-memory size of a real GPU launch as a \case (\S\ref{sec:design:interface}). 
We then construct
comparisons against an independent implementation, a naive reference
drafted by an LLM-based agent under our direction, or another
configuration of the same kernel. The last comparison uses the same
inputs and changes a configuration expected to preserve the
operator's semantics.

\heading{Counting.}
We review refuted and unresolved output comparisons alongside
engine diagnostics. Of the \allbugs bugs, 35 are exposed
by race or memory diagnostics, 41 by output refutations, and six by
manual inspection of cases that remain \vunknown.
An \vunknown result alone does not establish a bug.

The study excludes duplicates and previously unknown findings not filed
upstream. All \newbugs new bugs have an upstream
issue or pull request.
Table~\ref{tab:bugs-per-library} reports their
distribution. These counts refer to distinct bugs, not distinct buggy
kernels or the number of refuted kernels in \S\ref{sec:eval:coverage}.
\confbugs of the \newbugs new bugs are
confirmed: 29 are fixed upstream, four are acknowledged by maintainers,
and eight are independently validated (six with a proposed fix and
two with a public endorsement).

\begin{table}[t]
\centering
\footnotesize
\caption{Bugs in the study by library. Found includes previously
unknown bugs and rediscoveries; new denotes previously unknown bugs
filed upstream; confirmed denotes new bugs fixed or acknowledged by
maintainers, or validated by an independent contributor.}
\label{tab:bugs-per-library}
\begin{tabular}{lrrr}
\toprule
Library & Found & New & Confirmed \\
\midrule
FlagGems & 25 & 23 & 5 \\
flash-linear-attention & 11 & 11 & 11 \\
Attorch & 6 & 6 & 6 \\
CuPy & 6 & 6 & 1 \\
Triton-distributed & 4 & 3 & 1 \\
torchao & 4 & 4 & 3 \\
Liger-Kernel & 3 & 2 & 2 \\
vLLM & 2 & 0 & 0 \\
15 others & 21 & 21 & 12 \\
\midrule
\textbf{Total} & \textbf{\allbugs} & \textbf{\newbugs} & \textbf{\confbugs} \\
\bottomrule
\end{tabular}
\end{table}

\subsection{Comparison with Other Methods}
\label{sec:eval:compare}

We compare \sys with differential testing and two tools:
Volta~\cite{volta}, used as shipped, and
\texttt{compute-sanitizer} from NVIDIA~\cite{nvidia2025computesanitizer}.
The comparison uses the set of \allbugs bugs found by \sys.

\begin{figure}[t]
\centering
\includegraphics[width=\columnwidth]{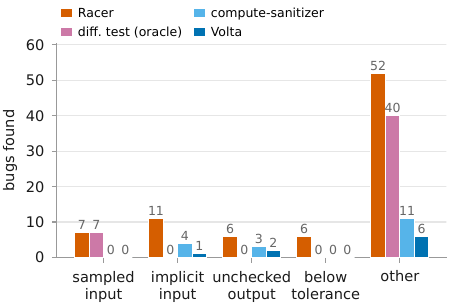}
\caption{Comparison of \sys, oracle differential-testing, \texttt{compute-sanitizer}, and Volta on the 82-bug set.}
\label{fig:compare-bars}
\end{figure}

We assess an oracle differential tester with unlimited compute:
given a triggering launch configuration and an independent reference,
it tests every possible tensor input using a tolerance-based output
comparison. This is a per-bug assessment, not a measured testing campaign.
Sanitizer and Volta results come from recorded runs.
Figure~\ref{fig:compare-bars} groups the results using the categories
of \S\ref{sec:bug}.

\heading{Differential testing.}
Under exhaustive input enumeration, differential testing can detect
all seven sampled-input bugs: it reaches the exact boundary values
and special inputs that random generators can miss. Together with the
40 bugs in Other identified by the audit, this gives 47 of \allbugs
bugs. Exhausting tensor inputs does not by itself ensure variation of
prior memory state or thread schedules, inspection of writes outside
the output tensor, or detection of differences within the tolerance.
The remaining assessments retain the audit's judgments under these
conditions; they do not establish that no concrete test can expose
those bugs.

\heading{\texttt{compute-sanitizer}.}
The recorded runs detect \computesanitizerbugs bugs:
four in implicit input, three in unchecked output, and eleven in Other.
Seven of the detections in Other are read-only overruns. These
results illustrate the value of checking memory accesses in addition
to comparing output values. None of these runs reports the four race bugs; 
this result is specific to the configurations and executions evaluated.

\heading{Volta.}
We run the shipped Volta baseline and it detects \voltabugs bugs:
one in implicit input, two in unchecked output, and six in Other.
It detects none of the sampled-input or below-tolerance bugs,
and none of the four races.

The recorded failures identify concrete barriers to applying Volta
to these kernels. In 41 cases, unsupported instructions prevent
execution. One example is the Triton-distributed
reduction in \S\ref{sec:bug}: Volta stops at an atomic instruction
before it can expose the uninitialized-memory bug.

Another limitation arises when the two implementations divide work
differently across CTAs. In the FlagGems MSE-backward comparison,
the tested kernel writes four output cells in CTA 0, while the
reference writes one cell per CTA across four CTAs. Because Volta
compares only CTA 0 on each side, it encounters different sets of
written cells and cannot compare their values. \sys executes both
grids in full, allowing it to compare corresponding output cells
regardless of which CTA writes them. Three study cases encounter
this limitation in Volta. Four others require comparing programs
with multiple launches, which Volta does not support.

\subsection{Coverage}
\label{sec:eval:coverage}

We assess \sys's coverage by measuring how many kernels it can
execute and how often it reaches an equivalence decision.
Each kernel is evaluated through several checking instances, each
specifying a launch configuration. Their outcomes can differ: a
kernel may be proved equivalent at one configuration, remain
unresolved at another, and be refuted at a third.
Table~\ref{tab:coverage} summarizes the evidence across these instances.
We group kernels with an \vneq or \vunknown result, or a race or
memory diagnostic, as \emph{bug candidates}. This group takes priority
over a proof at another configuration. An \vunknown result alone
does not establish a bug; these candidates require triage.

\begin{table}[t]
\centering
\footnotesize
\caption{Coverage over \numkernels kernels, summarized across their
checking instances. Each kernel appears in exactly one row.}
\label{tab:coverage}
\begin{tabular}{lrr}
\toprule
Kernel summary & Kernels & Share \\
\midrule
\veq & 1{,}698 & 70.5\% \\
Bug candidates & 197 & 8.2\% \\
Timeout, no decision & 14 & 0.6\% \\
Execution refused & 499 & 20.7\% \\
\midrule
\textbf{Total} & \textbf{\numkernels} & \textbf{100.0\%} \\
\bottomrule
\end{tabular}
\end{table}

The \veq row contains 1{,}698 kernels with at least one proof and
no recorded bug-candidate outcome. Each proof applies only to the
outputs and index assignments checked at that configuration.
The 197 candidates include 77 kernels with refutations, 23 with
\vunknown results, and 110 with race or memory diagnostics; these
counts overlap. A kernel appears in this row even if another
configuration is proved equivalent. Timeouts can also coexist with
other outcomes: 132 of the 146 kernels with timed-out instances have
a proof or refutation elsewhere; the remaining 14 appear separately.

\heading{Extending coverage.}
The remaining 499 kernels (20.7\%) fall in the execution-refusal row.
These refusals include unsupported integer operations, addresses or
control flow that depend on symbolic data, synchronization patterns
beyond the current model, and execution limits (\S\ref{sec:impl}).
We leave extending this coverage to future work.

\heading{Resolving inconclusive results.}
Of the \allbugs bugs, 35 produce race or memory diagnostics.
Among the remaining bugs, CAS cancellation alone detects one:
a constant bias in FlagGems' running-statistics update leaves a
nonzero constant after cancellation. For the other output comparisons,
symbols remain in the difference, so cancellation leaves these comparisons unresolved.
Exact numeric refutation resolves 36 of these cases, and the Z3
second prover resolves four more. Together, these techniques resolve
40 of the 46 cases (87.0\%) left inconclusive by cancellation.
Including the diagnostics and constant-bias refutation, \sys
automatically exposes 76 of the \allbugs bugs (92.7\%). The remaining
six retain \vunknown verdicts; we identify their bugs by manually
inspecting the kernel code and symbolic expressions.

\subsection{Shape Reduction vs.\ Whole-Grid Checking}
\label{sec:eval:scale}

We compare deciding each output cell independently (whole-grid
checking) with reusing proofs through shape reduction
(\S\ref{sec:impl:scale}). We measure end-to-end wall-clock time,
including symbolic execution, deciding, and other engine stages.

Our sample contains 50 kernels: 40 with \veq verdicts at their
captured shapes and ten with bugs. We run each at its captured
shape ($\times 1$) and at $\times 2$ and $\times 4$, scaling one
selected dimension while keeping the others fixed. This gives
150 kernel--configuration pairs. 

Shape reduction preserves every proof and refutation returned by
whole-grid checking: the methods agree on all 148
kernel--configuration pairs that both decide. Both decide all
50 kernels at $\times 1$ and $\times 2$, and both refute all ten
bugs at every shape. At $\times 4$, whole-grid checking decides
48 kernels and shape reduction decides 49; one kernel times out
under both methods. Thus, there are no contradictory verdicts,
and shape reduction proves one additional configuration within
the budget.

Shape reduction substantially reduces the cost of repeated proofs.
For the linear-attention backward kernel
\texttt{chunk\_ttt\_}\allowbreak\texttt{linear\_bwd\_dv}, it reduces end-to-end time
from 204.9\,s to 17.4\,s at $\times 1$ and from 415.8\,s to
34.0\,s at $\times 2$, speedups of $11.8\times$ and $12.2\times$.
At $\times 4$, whole-grid checking exceeds 600\,s without a
proof, whereas shape reduction proves equivalence in 70.7\,s.
It uses only 16 representative proofs for 2,048, 4,096, and
8,192 output cells, respectively. A separate configuration of
this kernel, scaling a different shape dimension, shows the same
benefit: whole-grid checking again exceeds 600\,s, while shape
reduction requires only 66.7\,s.

Across the 40 equivalent kernels at $\times 1$, shape reduction
lowers mean end-to-end time from 14.5\,s to 6.0\,s, an aggregate
speedup of $2.4\times$. Aggregate speedup is the ratio of summed
end-to-end times over the same completed comparisons.
Table~\ref{tab:reduce-cost} follows the fixed set of 38 equivalent
kernels completed by both methods at every shape. It excludes
both kernels with timeouts from all three columns. Shape reduction
reduces their mean end-to-end cost by $1.72\times$, $1.55\times$,
and $1.55\times$, respectively. At $\times 4$, mean deciding time
falls from 18.7\,s to 2.1\,s, while symbolic execution still takes
about 28\,s under either method. Symbolic execution therefore 
limits the end-to-end speedup even when proof reuse substantially 
reduces deciding time.

\begin{table}[t]
\centering
\small
\caption{Mean end-to-end time in seconds for the same 38 equivalent
kernels completed by both methods at all three shapes. Kernels
with a timeout under either method are excluded at every shape.}
\label{tab:reduce-cost}
\begin{tabular}{lrrr}
\toprule
Shape multiplier & $\times 1$ & $\times 2$ & $\times 4$ \\
\midrule
Whole-grid checking & 9.5 & 22.2 & 47.0 \\
Shape reduction & 5.5 & 14.3 & 30.3 \\
\midrule
End-to-end speedup & $1.72\times$ & $1.55\times$ & $1.55\times$ \\
\bottomrule
\end{tabular}
\end{table}
\section{Related Work}
\label{sec:relwork}

\heading{Testing and property checking for GPU kernels.}
Existing tools expose functional errors through testing or check specific
kernel properties.
NNSmith~\cite{liu2023nnsmith}, Tzer~\cite{liu2022tzer}, and
NeuRI~\cite{liu2023neuri} generate or mutate tensor programs to expose
compiler bugs. Differential output comparisons can reveal functional
errors, subject to the testing limitations discussed in \S\ref{sec:motivation}.
cuFuzz~\cite{ziad2026cufuzz} uses coverage-guided fuzzing of CUDA
applications with sanitizer integration. NVIDIA's
\texttt{compute-}\allowbreak\texttt{sanitizer}~\cite{nvidia2025computesanitizer}
checks memory accesses and synchronization during execution.
Race detectors include Barracuda~\cite{eizenberg2017barracuda},
CURD~\cite{peng2018curd}, and ScoRD~\cite{kamath2020scord};
Simulee~\cite{wu2020simulee} models
memory accesses to detect synchronization bugs.
PUG~\cite{li2010pug}, GPUVerify~\cite{betts2012gpuverify,betts2015gpuverify},
GKLEE~\cite{li2012gklee}, and ESBMC-GPU~\cite{pereira2016esbmc}
use symbolic execution or formal verification to check properties such as
race freedom and synchronization correctness within their supported
execution models and analysis bounds;
PUG and GKLEE also check functional assertions.
Model execution can provide input constraints for checking kernel memory safety.
M2K~\cite{he2026m2k} checks CUDA memory safety under argument constraints
inferred from model execution, using symbolic tensor sizes and thread identifiers to cover
variable shapes and launch configurations.
\sys instead compares the functions computed by two programs at declared
tensor shapes and launch configurations, an equivalence question that
memory and synchronization checks alone do not resolve.

\heading{Program equivalence and translation validation.}
Comparing program behavior provides a basis for validating compiler
transformations and alternative implementations.
Translation validation checks the correctness of individual compiler
transformations~\cite{pnueli1998translation,necula2000translation};
Alive2~\cite{lopes2021alive2} provides bounded validation for LLVM.
Regression verification, SymDiff, and product programs reason about
relationships between programs~\cite{godlin2009regression,lahiri2012symdiff,barthe2011product};
KLEE~\cite{cadar2008klee} uses symbolic execution to cross-check
implementations along explored paths.
GPU-kernel equivalence additionally requires modeling communication and
synchronization among threads whose work assignments may differ between
implementations (\S\ref{sec:background:symbolic}).

\heading{Tensor-program verification.}
Tensor-program verifiers reason about computations expressed in tensor
graphs or specialized languages.
At the tensor-graph level, TASO~\cite{jia2019taso} and
TensorRight~\cite{arora2025tensorright} prove rewrites correct under their
operator semantics.
TrainVerify~\cite{lu2025trainverify} checks a distributed training plan
against its single-device specification. Its shape reduction shrinks
tensor shapes while preserving the equivalence question over the modeled
operators. \sys instead executes the compiled kernels and reuses an
output-cell proof only when a validated shared renaming relates both
programs' expression pairs (\S\ref{sec:impl:scale}).
Mirage~\cite{wu2025mirage} uses probabilistic equivalence verification
with formal guarantees over its representation of tensor programs at the
kernel, thread-block, and thread levels.
Ave~\cite{mai2026ave} guides kernel optimization by checking data-flow
invariants expressed as symbolic tags and assertions in its tensor language,
and tests candidate kernels for functional correctness.
\sys compares symbolic output expressions derived from compiled PTX,
modeling the memory accesses and synchronization that implement the computation.

\heading{GPU-kernel equivalence checking.}
Functional proofs can also connect CUDA kernels to high-level specifications.
ProofWright~\cite{chatterjee2025proofwright} uses LLMs to generate VerCors
annotations for CUDA memory and thread safety, and combines Rocq proofs
with functional annotations to relate kernels to PyTorch specifications.
Its evaluated functional proofs cover a subclass of element-wise kernels;
the translation from Rocq specifications to VerCors annotations is
manually validated.

Volta~\cite{volta} is the closest prior work, sharing \sys's approach of
symbolically executing PTX over exact reals to check equivalence.
It provides an equality decision procedure for its
supported expression class. It also checks concurrency properties under
its structured-CTA model and rejects programs outside that model.
\sys compares complete output tensors across all CTAs and successive
launches, allowing the two implementations to partition work
differently. It also supports additional production PTX constructs
and reuses output-cell proofs through shape reduction.
 
\section*{Acknowledgments}
ChatGPT and Claude assisted with drafting and revising the manuscript. 
LLM-based agents also assisted with constructing reference kernels
and investigating candidate bugs, as described in Section~\ref{sec:eval:hunt}.
The authors are responsible for the paper's content and conclusions.

\bibliographystyle{ACM-Reference-Format}
\bibliography{references}

\clearpage
\appendix
\section{Shape Reduction and CTA Scheduling Algorithms}
\label{sec:appendix:algorithms}

Algorithm~\ref{alg:shape} gives the candidate-renaming procedure used
by shape reduction (\S\ref{sec:impl:scale}). The candidates must pass
structural validation on both programs' expressions before a proof can
be reused. Algorithm~\ref{alg:scheduler} gives the CTA scheduler
described in \S\ref{sec:impl:threads}; each thread's memory accesses
are checked by the memory model as it executes.

\begin{algorithm}[H]
\footnotesize
\caption{Candidate renamings for shape reduction (\S\ref{sec:impl:scale}):
candidate re-indexings $\tau$ from a proven cell $r$ to a cell $c$, read
off their input footprints alone; $\rho_T(c)$ is the set of flattened indices
in input tensor $T$ referenced by either program's expression at $c$.
Cell $c$ reuses $r$'s equality proof only if
$e^a_c = \tau(e^a_r)$ and $e^b_c = \tau(e^b_r)$
structurally.}
\label{alg:shape}
\begin{algorithmic}[1]
\Function{Propose}{$r$, $c$}
  \If{$\rho(r)$ and $\rho(c)$ name different tensors} \Return $\emptyset$
  \EndIf
  \ForAll{input tensors $T$}
    \State $d_T \gets \min \rho_T(c) - \min \rho_T(r)$
    \Comment{shift of the lowest index}
    \If{$\rho_T(r) + d_T \neq \rho_T(c)$} \Return $\emptyset$
    \Comment{not a translation}
    \EndIf
  \EndFor
  \State $\tau_1 \gets \big(T[x] \mapsto T[x + d_T]\big)$ for every $T$
  \Comment{translation}
  \If{$r \in \rho_T(r)$ and $c \in \rho_T(c)$ for some $T$}
    \Comment{row softmax}
    \State $\tau_2 \gets \tau_1$ followed by $T[r + d_T] \leftrightarrow T[c]$
    \State \Return $\{\tau_1, \tau_2\}$
  \EndIf
  \State \Return $\{\tau_1\}$
\EndFunction
\end{algorithmic}
\end{algorithm}
\begin{algorithm}[H]
\footnotesize
\caption{The CTA scheduler.}
\label{alg:scheduler}
\begin{algorithmic}[1]
\Statex \textbf{Notation.}
\Statex \hspace{\algorithmicindent} $T$: the threads of one CTA;\quad
  $Q \subseteq T$: the threads ready to run.
\Statex \hspace{\algorithmicindent} $S$: the synchronization points of the
  kernel.
\Statex \hspace{\algorithmicindent} $P(s) \subseteq T$: the threads that
  $s \in S$ synchronizes.
\Statex \hspace{\algorithmicindent} $W[s] \subseteq P(s)$: the threads
  waiting at $s$.
\Statex \hspace{\algorithmicindent} $M$: the CTA's memory model.
\Statex
\State $Q \gets T$;\quad $W[s] \gets \emptyset$ for every $s \in S$
\While{$Q \neq \emptyset$}
  \State pop a thread $t$ from $Q$
  \State run $t$ symbolically on $M$, checking every access, until it
    exits or reaches some $s \in S$
  \If{$t$ reached $s$}
    \State $W[s] \gets W[s] \cup \{t\}$
    \If{$W[s] = P(s)$} \Comment{every thread $s$ waits for has met}
      \If{$P(s) = T$} $M.\mathrm{endEpoch}()$
        \Comment{publish writes}
      \EndIf
      \State resolve $s$;\quad $Q \gets Q \cup W[s]$;\quad
        $W[s] \gets \emptyset$
    \EndIf
  \EndIf
\EndWhile
\If{$W[s] \neq \emptyset$ for some $s \in S$}
  \State \textbf{refuse}: divergence \Comment{some threads never met}
\EndIf
\end{algorithmic}
\end{algorithm}
  
\end{document}